\documentclass[preprint,aps,tightenlines,superscriptaddress,floatfix,
showpacs,nofootinbib]{revtex4}
\usepackage[final]{graphicx}
\usepackage{epsfig}
\usepackage{bm}

\newcommand{\boldtau}{\mbox{\boldmath $\tau$}}
\newcommand{\boldphi}{\mbox{\boldmath $\phi$}}
\newcommand{\boldpi}{\mbox{\boldmath $\pi$}}

\newcommand{\boldsigma}{\mbox{\boldmath $\sigma$}}
\newcommand{\he}{\mbox{$^3{\rm He}$}}
\begin{document}

\title{Positive pion absorption on $^3$He using modern 
trinucleon wave functions}

\author{S. Schneider} \author{J. Haidenbauer} \author{C. Hanhart}
\affiliation{Institut f\"ur Kernphysik, Forschungszentrum
J\"ulich, D-52425 J\"ulich, Germany}
\author{J. A. Niskanen}
\affiliation{Department of Physical Sciences, PO Box 64,
FIN-00014 University of Helsinki, Finland}


\begin{abstract}
We study pion absorption on $^3{\rm He}$ employing 
trinucleon wave functions calculated from modern realistic 
$NN$ interactions (Paris, CD Bonn). 
Even though the use of the new wave
functions leads to a significant improvement over older calculations
with regard to both cross section and polarization data, there are hints
that polarization data with quasifree kinematics cannot be
described by just two-nucleon absorption mechanisms.
\end{abstract}


\maketitle

\section{Introduction}
One hope in building the so called meson factories towards the
end of the 70's was to use mesons, in these facilities pions,
as probes of nuclear wave functions and nuclear structure
at short distances \cite{lock}.
However, on the theoretical side it soon turned out that
meson interactions even with the two-nucleon systems were
quite a challenge and most work concerned these \cite{garci}.
Pion production physics obtained a new surge with the advent of
a new generation of accelerators at IUCF, Celsius and COSY
with a very high energy resolution making possible accurate
measurements at meson thresholds \cite{review}.
New and even unexpected results also created renewed
theoretical activity, concentrated still mainly on two-nucleon
meson production at threshold and  also at higher energies to
understand some puzzles e.g. in $pp\rightarrow pp\pi^0$
threshold production \cite{meyer}.
Nevertheless, there has emerged a general consensus
of a fair understanding of at least the main mechanisms
in the two-nucleon system, although some problems still
remain -- within the conventional (meson-exchange) approach \cite{MMM}
as well as in the chiral perturbation treatment of pion
production and absorption \cite{HHH}.

New experiments are also performed
or in progress on meson production in
few-nucleon systems as in $pd\rightarrow$ $^3$He$\,\pi^0$ or
$pd\rightarrow$ $^3$H$\,\pi^+$ \cite{pdhepi} as well as
corresponding $\eta$ meson production experiments \cite{pdheeta}.
However, theoretical efforts in this direction
with three-nucleon dynamics are very scarce \cite{Canton,Green}
and the situation is much less satisfactory as compared with
the two-nucleon case. Nevertheless,
pionic inelasticities in 3- or 4-nucleon
systems should be the necessary bridge towards understanding them
in nuclei and potentially using them as a probe in many-body nuclear
physics and for possible effects of nuclear medium on hadrons and
their interactions. One may also note that in these phenomena some
reaction channels are actually only accessible in absorption.

At the above mentioned new facilities pion absorption experiments
are unlikely due to their low intensities. However, absorption is
closely related to production reactions and should be understood in
parallel. Furthermore, it may be argued that some absorption processes
might be
easier to approach theoretically than production. One such process
could be quasifree absorption on a pair of nucleons in $^3{\rm He}$
(or in triton). This is the inverse of two-nucleon pion production
in the presence of a (hopefully) inactive spectator. Here the initial
state nuclear wave function is known, in principle, exactly from
Faddeev calculations and the final state pair is similar to those
treated in two-nucleon reactions. Success in this simplest case
might open the door to modelling (with explicit inclusion of the
spectator) three-nucleon absorption (where data from PSI \cite{weyer}
are available)
and the break-up of $^3{\rm He}$ into a deuteron and a proton --
the inverse of the above referred production reactions.

Experimental cross sections of quasifree two-nucleon absorption
of pions on helium isotopes have been obtained from the meson
factories of LAMPF \cite{lampf}, TRIUMF \cite{triumf} and
PSI \cite{weber}, but scarce data exist also for the polarization
of outcoming fast protons \cite{maytalbeck,aclander}.
These are obtained at so called conjugate angles
corresponding to kinematics where it is believed that
the spectator is not an active
participant and does not absorb momentum from the pion. Then the
spectator remains essentially at rest retaining only its Fermi
momentum. In Ref. \cite{triumf} one sees at these angles a massive
peaking of the cross section, over an order of magnitude higher
than for nonconjugate angles, as a function of the proton energy.
The width of this peak may be accounted for with the Fermi motion.
The quasifree nature (the spectator having essentially the momentum
distribution of the bound state) is even more convincingly established
in the kinematically complete experiments of Ref. \cite{weber}.
Cross sections for positive and negative pion absorption on tritium
were obtained in Ref. \cite{salvisberg}. Overall, this gives
a good wealth of data to determine absorption on different nucleon
pairs with different isospins in a simple nuclear environment.
Also heavier nuclei have been investigated in related contexts
\cite{altman}.

Theoretical work is of old vintage, the most recent serious work
probably being in Refs. \cite{quasid,quasids} for positive pions and
\cite{piminus,impact} for negative pions,
and \cite{kiang} for branching ratios in stopped $\pi^-$
absorption.  The angular shapes of the cross sections could
be well explained and, roughly, also absolute magnitudes.
In fact, for positive pions the shapes don't differ much
from $pp\rightarrow d\pi^+$ in theory or experiment. However,
in absorption of positive pions on $^3{\rm He}$
or $^4$He the polarization of the outcoming protons was found
to be in qualitative disagreement with the simple theory
employed \cite{maytalbeck,aclander}, which neglected the
effect of the spectator and used phenomenological range-corrected
deuteron wave functions to describe the active pair as a
quasideuteron. The
measurements were performed at 120 and 250 MeV and it was
possible to reproduce the data qualitatively  - however, only
when applying different models for the two energies, and not with
the same model for both energies.

In a recent paper \cite{param} a simple parametrization was
presented that approximates analytically exact three-nucleon
bound state wave functions resulting from Faddeev calculations
based on realistic nucleon-nucleon ($NN$) interactions.
This parametrization is similar in philosophy to that
of Ref. \cite{hgs} but deviates from it in two important ways.
Firstly, it releases its single term separability in
the two relative momenta $p$ and $q$ of the pair and the
spectator (or the corresponding coordinates $r$ and $\rho$).
This gives more freedom for reproducing the
behaviour of the wave function better when both momenta are large --
as one would expect, for example, that one particle which is far off
shell would influence the others.
In contrast, the parametrization of Ref. \cite{hgs}
treats the dependence of the wave function on the two momenta $p$
and $q$ as being totally independent of each other.
There are actually significant differences between the wave
functions at momenta relevant for mesonic inelasticities \cite{param}.
It is interesting to see what impact these may have to physically
observable quantities, in particular, whether the pair-spectator
correlation could correct the above mentioned energy dependent
discrepancy seen in the pion absorption reactions.

A second difference is that,
instead of parametrizing the single Faddeev amplitudes only,
corresponding to different permutations of the three nucleons,
as done
in Ref. \cite{hgs}, we parametrized directly partial wave
projections of the total antisymmetrized wave function. Also
this expansion was seen to be well convergent and was applied in
calculations of low-momentum quantities such as the probabilities
of the trinucleon wave function components
and the $\pi^3$He scattering length in Ref. \cite{param}.

In the present paper we apply this parametrization to study
quasi-two-body absorption of pions on $^3$He. Thereby,
the aim is two-fold. Firstly, we want to test the reliability
and convergence of our parametrization of the three-nucleon bound state
wave function in calculations of observables involving higher momenta.
Secondly, we want to see how one can fare with such improved wave
functions in this specific reaction physically,
without explicit participation of the spectator nucleon.
In the following section we shortly outline the most essential
features of the parametrization and provide some details of the
ingredients and technical aspects of our calculation of pion absorption,
while the third section deals with the actual results of our
pion absorption calculation.
The paper ends with some concluding remarks.

\section{Formalism}
\subsection{Faddeev amplitudes in \he}
Aiming at extreme simplicity the model of Ref. \cite{quasid}
considered quasifree absorption of positive pions as simply
absorption on a quasideuteron with a wave function more
compressed than the free deuteron (because the binding energy
is larger) and with kinematics
compatible with 10 MeV more binding than in the normal
deuteron (5 MeV for the actual binding energy difference
{\it plus}
5 MeV for the average kinetic energy of the spectator from
its momentum distribution). A similar approach was adopted
also later for negative pion absorption on a singlet proton
pair \cite{piminus,impact} and actually was able to explain
such features of the differential absorption cross section as
the asymmetry about $90^\circ$ and also of the analyzing
power in the closely related process $\vec p n \rightarrow
(pp)_{S{\rm -wave}}\pi^-$.

The trinucleon wave functions adopted were basically of two kinds.
Initially, phenomenological functions based on a range-modified
deuteron wave function following an old idea of Ref.
\cite{hadji} or on a calculated correlation function
\cite{friar} were used in Ref. \cite{quasid}. Later
also Faddeev pair wave functions $v(r)$ from the
separable form
$\psi^\nu ({ r},{\rho}) = v^\nu({r})w^\nu({\rho})$
parametrized by Hajduk {\it et al.} \cite{hgs} were
used in Refs. \cite{piminus,impact}, where
\begin{equation}
\psi^\nu (r_{ij},\rho_k)
= \langle r_{12}\rho_3\nu_{12} |\psi [(12)3]\rangle
= \langle r_{23}\rho_1\nu_{23} |\psi [(23)1]\rangle
= \langle r_{31}\rho_2\nu_{31} |\psi [(31)2]\rangle
\label{single}
\end{equation}
and the total antisymmetric wave function is
\begin{equation}
|\Psi \rangle = |\psi [(12)3]\rangle +
|\psi [(23)1]\rangle + |\psi [(31)2]\rangle . \label{total}
\end{equation}
However, these calculations used for absorption on each
pair $ij$ only the wave function component above with the
particular permutation $(ij)k$ and considered only the
corresponding Jacobian coordinate $r_{ij}$ in the absorption
process. With a single term separable parametrization \cite{hgs}
or with a completely phenomenological pair wave function this
left the role of the spectator to a mere normalization integral.
Plausibly, the use of the square root of the correlation
function as the pair wave function may take the other two terms
in Eq.~(\ref{total}) effectively into account to some extent.
Evidently, this issue will now be addressed more explicitly with
the new wave functions.

In the above functions the index $\nu$ labels the partial wave
structure of the three nucleons. In the following calculations
we only consider the states
with zero spectator orbital angular momentum, so that this
index trivially just symbolizes the quantum numbers of
the pair wave functions, in the singlet spin state
$^1S_0$ and in the triplet $^3S_1$ or $^3D_1$. The two additional
states with the spectator angular momentum 2 considered in
Refs. \cite{param,hgs} have much less weight and
are assumed to be of little 
importance for the present kinematics where the spectator remains
essentially at rest.  

In Ref. \cite{param} a considerably different parameterization
was given for the wave functions. Firstly, the simple separability
used in Ref. \cite{hgs} was generalized to more terms of
separable form with a systematic improvement in the approximation.
The structure of the wave function remained basically simple
but allowed  correlation between the momenta
or the corresponding coordinates, which is not present in the
simple product ansatz.  Physically one
might expect that, if in a bound three-body system either the
spectator or the pair is far off-shell, then it would be less
likely to find also the other far off-shell. A parameterization
as a sum of two products was seen to offer sufficient
freedom and to allow a reasonable fit in the sense that inclusion of
a third term did not have much effect. It is worth noting that
at large momenta, relevant to meson production and absorption,
the inclusion of the second term changed the wave function
significantly.

The second essential difference is that in Ref. \cite{param}
a parameterization for the fully antisymmetrized wave function
was provided, and not only for its
individual Faddeev amplitudes as in Ref. \cite{hgs}. This is a
nontrivial extension, including also the two other amplitudes
of Eq.~(\ref{total}) in
the projection on angular momentum eigenstates, and has
the advantage that all permutations enter automatically into
the calculation of each pair absorption but still with simple
wave functions for a given coordinate pair.
E. g. if the form of Eq.~(\ref{total}) is used in absorption
on the pair 12, the first term is simple,
but in the other terms the "proper" simple pair coordinate
would be ${\bf r}_{23} =-1/2\, {\bf r}_{12}-{\bm\rho}_3$
or ${\bf r}_{31} =-1/2\, {\bf r}_{12}+{\bm\rho}_3$. These terms
would be quite complicated functions of
the coordinates ${\bf r}_{12}$ and ${\bm\rho}_3$.
However, once the full antisymmetric wave function is parametrized
directly in terms of ${\bf r}_{12}$ and ${\bm\rho}_3$,
the calculation is greatly simplified. The choice of the pair
does not matter,  since physically
absorption on any pair should give the same result, anyhow.

In practice the full antisymmetric Faddeev wave
function (calculated using the CD Bonn \cite{bonn} and Paris 
\cite{paris} potentials) was expressed as a product of functions of
the pair and spectator momenta ${p}$ and ${q}$, where each
function is given by expansions in terms of Lorentz functions
\begin{equation}
\tilde v_1^\nu (p) = \sum_i \frac{a_i^\nu  }{p^2+(m_i^\nu )^2},\qquad
\tilde w_1^\nu (q) = \sum_i \frac{b_i^\nu }{q^2+(M_i^\nu )^2}
\end{equation}
for the five most important Faddeev amplitudes.
In the coordinate representation these functions will transform
into Yukawa functions and (for $D$ waves) their derivatives,
\begin{equation}
v_1^\nu (r) = \sqrt{\frac \pi 2}\sum_i a_i^\nu   e^{-m_i^\nu r}\quad
{\rm or}\quad
v_1^\nu (r) = \sqrt{\frac \pi 2}\sum_i a_i^\nu   e^{-m_i^\nu r}
\left( 1+\frac{3}{m_i^\nu r}+\frac{3}{(m_i^\nu )^2r^2}\right)\, ,
\end{equation}
with similar expressions for the (spectator) $\rho$ dependence.
The denominator $r$ is cancelled against the volume element.

\begin{figure}[t!]
\epsfig{figure=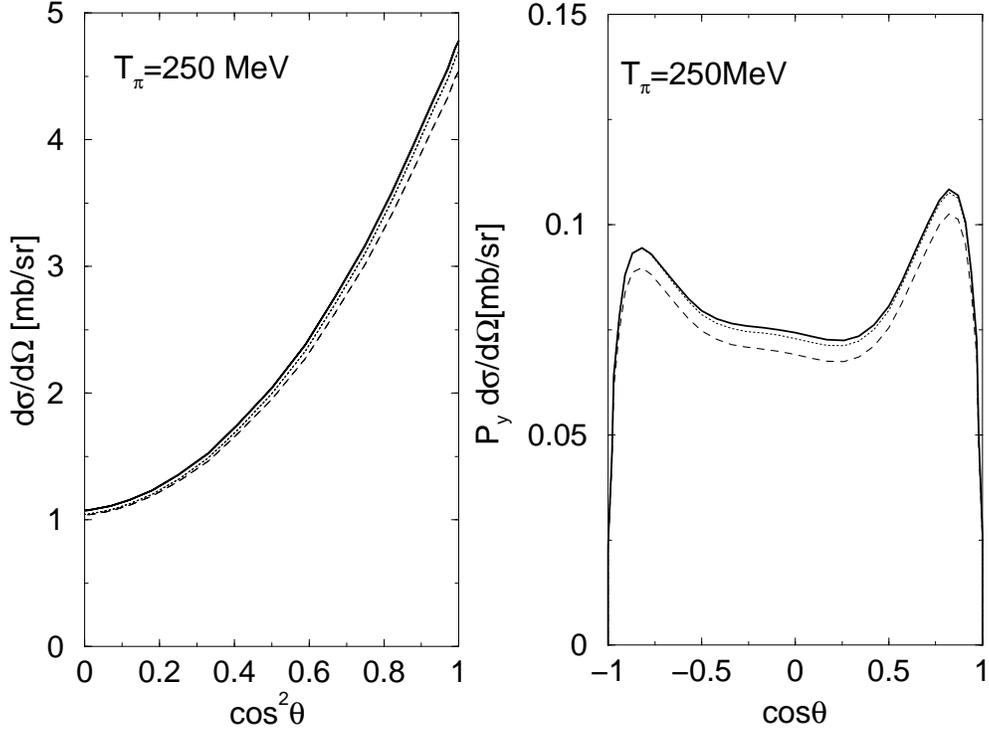,width=10cm,angle=-90}
\caption{The differential absorption cross section and
its "asymmetry" $P_y\, d\sigma /d\Omega$ at $T_\pi$ = 250 MeV
using different fits to the CD Bonn trinucleon wave function.
Dashed: single-term separable fit; solid: two-term fit; dotted:
three-term fit.\label{convcross}}
\end{figure}

Up to this point the procedure would have been equivalent
to Ref. \cite{hgs}, except that the fit was performed to the exact
total (antisymmetrized) wave functions. However, in Ref. \cite{param}
another similar product term $v_2^\nu (p) w_2^\nu (q) $
was added in order to improve the quality of the analytical
representation of the wavefunction. Thus the
wave function was presented as
\begin{equation}
\Psi^{\nu}(p,q)=v^{\nu}_1(p)w^{\nu}_1(q)+v^{\nu}_2(p)w^{\nu}_2(q),
\label{twoterms}
\end{equation}
with the normalization
\begin{equation}
\sum_{\nu}\int_0^{\infty} dp\; dq\; p^2q^2 |\Psi^{\nu}(p,q)|^2=1.
\end{equation}
The parameters of the fit(s) were given in Ref. \cite{param}
and will not be repeated here, but the importance of the
additional freedom will be studied in the differential
cross section and polarization of the protons in quasifree
absorption on positive pions on quasideuterons. At this stage
it is worth remembering that for a specific low-momentum observable, the
${\pi^-}\,{^3}$He scattering length, the effect of the second term
in Eq.~(\ref{twoterms}) was seen to be only about the order of
1\% \cite{param}.


Before going into any details of the pion absorption mechanisms
we want to test the significance of nonseparability anticipated above
for physical reasons.
Therefore, we explore extreme momentum transfers
corresponding to the highest energy at which polarization data
in pion absorption are available, namely $T_\pi$ = 250 MeV. Our results for the
differential absorption cross section on a quasideuteron
and transverse polarization\footnote{To facilitate
better comparison this is multiplied by $d\sigma /d\Omega$.
In meson production this would correspond to the asymmetry of the
cross section of the two-nucleon reaction with a polarized beam.}
of an outcoming proton are shown in Fig. \ref{convcross} utilizing
a systematic expansion of the wave function
up to three separable terms, i.e. beyond Eq.~(\ref{twoterms}).
It can be seen that nonseparability does
play a significant role. However, it is encouraging to see that
just one additional term of products is sufficient
to account for it and that the expansion has
converged quite well already at the two-term level. In the
calculations presented in Fig. \ref{convcross}
the Faddeev wave functions from the CD Bonn
potential are used. We want to mention, however, that the
convergence features for those based on the Paris potential
were found to be the same.

\subsection{Absorption formalism}

\begin{figure}[t!]
\epsfig{figure=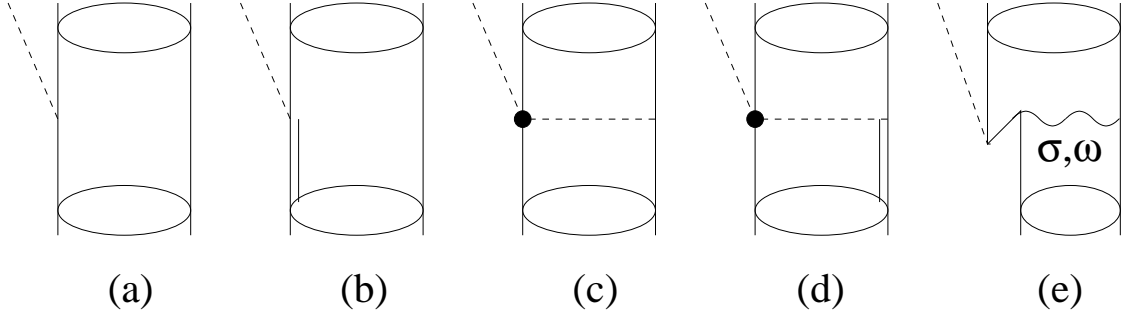,height=4cm,angle=0}
\caption{The mechanisms included in pion production (as well as in absorption)
on two nucleons:
a) direct production, b) "direct" production involving the $\Delta(1232)$
isobar, c) pion-nucleon $s$-wave rescattering, d) $s$-wave rescattering
of a pion originating from a $\Delta$, e) heavy meson exchange.
\label{mechs}}
\end{figure}

The mechanisms in pion absorption on two nucleons have been
discussed in detail elsewhere \cite{quasid,piminus,impact} and will not 
be repeated here in depth\footnote{Many of the existing calculations
are actually done for pion production, but time reversal is trivial.}.
They are depicted in Fig. \ref{mechs} for the time reversed reaction
corresponding to pion production. The first one (Fig.~\ref{mechs}a)
is the standard direct production due to the Galilean
invariant
$\pi N$ interaction arising from the pseudovector coupling (with
obvious notation)
\cite{ppdpi}
\begin{equation}
 H_{\pi NN} = {f \over m_{\pi}}\ \sum_i {\boldsigma}_i \cdot
\left\{  {\bf q}\  {\boldtau}_i \cdot {\boldphi} -
 {\omega_q \over 2M}
\left[ {\bf p}_i \ {\boldtau}_i\cdot {\boldphi} + {\boldtau}_i
 \cdot {\boldphi}\ {\bf p}_i \right] \right\} . \label{direct}
\end{equation}
Here the first term would give $p$-wave pions (relative to nucleon
$i$), while the second term when operating on the $NN$ wave function
facilitates also (mainly) $s$-wave production.
The direct production is generalized to include also $p$-wave
$\pi N$ rescattering (Fig. \ref{mechs}b) via the $\Delta(1232)$
resonance. Note that this contribution is treated on the same
footing as the direct production by generating first
the $\Delta N$ admixture by the coupled channels method
in the initial state. Subsequently the $\Delta$ decays by an
operator similar to Eq.~(\ref{direct})
(with the spin and isospin operators replaced by the
$\Delta N$ transition operators and the $\pi NN$ coupling constant
$f^2/4\pi = 0.076$ by the $\pi \Delta N$ coupling constant
$f^{*2} /4\pi = 0.35$ from the decay width of the $\Delta$).
This produces the well known prominent cross section peak at
pion energies around 150 MeV.

The $NN$ interaction of the high-energy nucleon pair is based
on the Reid soft core potential \cite{reid}. At high energies the
details of the potential are not expected to be very important.
Moreover, within a coupled channels treatment the $NN$ part must be
modified, anyway, to avoid doubly counting the attraction
generated by the coupling to $\Delta N$ intermediate states
\cite{ppdpi}.

At threshold, both production and absorption are, however, dominated by
$\pi N$ $s$-wave rescattering, Fig. \ref{mechs}c, with also a substantial
contribution from Fig. \ref{mechs}d. This rescattering is described by a
phenomenological $\pi N$ interaction
\begin{equation}
H_s = 4\pi \,\frac{\lambda_1}{m_\pi}\, \boldphi\cdot\boldphi +
4\pi\,\frac{\lambda_2}{m_\pi^2}\,
 \boldsigma\cdot\boldphi\times\boldpi\, ,
\end{equation}
where the parameters $\lambda_1$ and $\lambda_2$ depend on the
$\pi N$ on-shell momentum and are fitted
to pion-nucleon scattering data \cite{impact}. As far as the
relative importance in positive and negative pion absorption is
concerned one should note that, due to chiral invariance,
$\lambda_1$ is suppressed close to threshold
by a a factor of $m_\pi/M_N$ as compared with $\lambda_2$.
Indeed $\lambda_1$ is very small at threshold, but it
becomes comparable to $\lambda_2$ for pion momenta $q_\pi$
corresponding to $\eta = q_\pi /m_{\pi} \ge 0.5$. A monopole form
factor is included in the meson exchange interaction. The value of the
cut-off mass is crucial in fitting the analyzing power $A_y$
at some given energy. Its effect is small on the total cross section
except close to threshold.

It is known that the above mechanisms are not sufficient to
explain the size of the cross section of the reaction
$pp\rightarrow pp\pi^0$ \cite{meyer}.
The remaining strength
could be explained by short-range contributions from the $NN$
interaction to the axial charge of the two nucleons, most
importantly by exchanges of the $\sigma$ and $\omega$ mesons
as shown in Fig. \ref{mechs}e
\cite{lee,horo,comment}\footnote{As an alternative to this
heavy meson exchange mechanism also $\pi N$ off-shell 
rescattering has been proposed \cite{oset}. Reality may be a
combination of both \cite{hanhart}.}.
Consequently its effect was seen to be important also in
negative pion absorption on $^1S_0$ $pp$ pairs in $^3$He
\cite{impact}. In $pp\rightarrow d\pi^+$ and the inverse 
reaction (i.e. the present consideration with a quasideuteron) 
the effect of the
heavy meson exchange was seen to be much less important
\cite{comment}. However, in the present context the wave
functions are more condensed than in the deuteron and it
is of interest to include also this short range effect.
Further motivation for taking it into acount here
is provided by the possibility that the
active $pn$ pair can appear also in the $^1S_0$ state.

\begin{figure}[t!]
\epsfig{figure=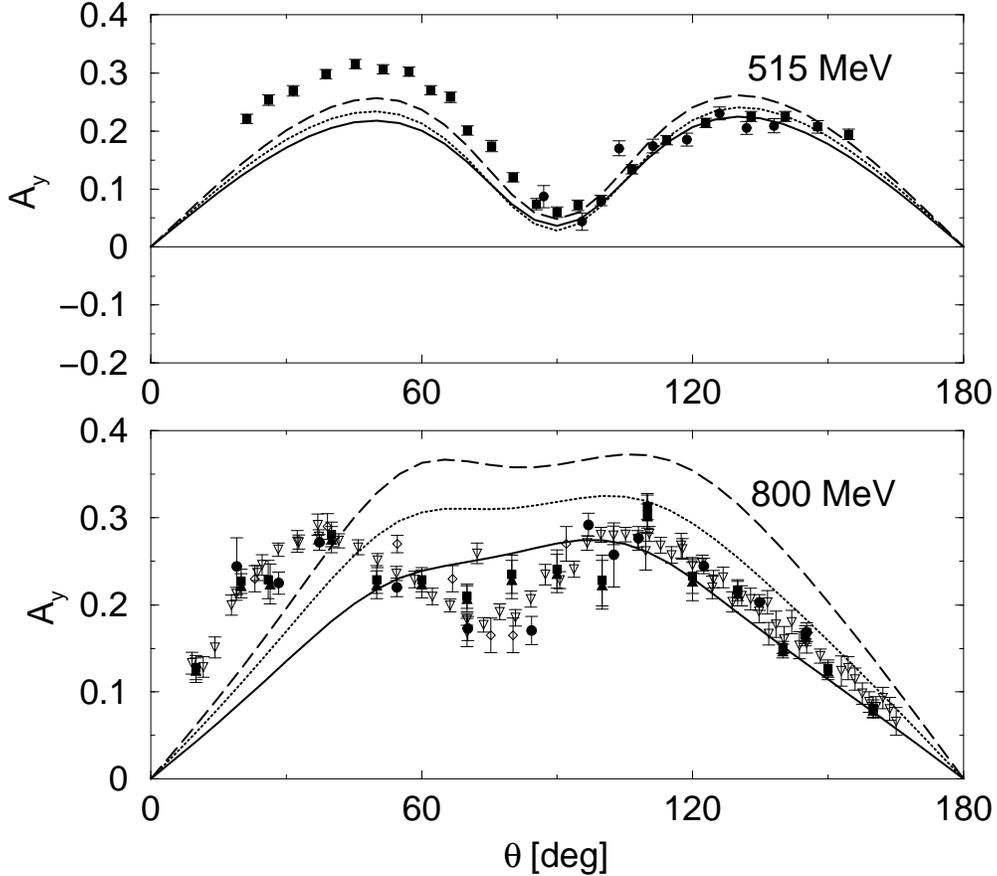,width=12cm,angle=-90}
\caption{The analyzing power $A_y$ in the reaction
$pp \rightarrow d\pi^+$ at two energies closely
corresponding to the $\pi^+\, ^3$He absorption energies of
Ref. \protect\cite{maytalbeck}. The different deuteron
wave functions used are: CD Bonn (solid curve); Paris (dotted curve);
Reid soft core (dashed curve). \label{ayppdpi}}
\end{figure}

As a starting point Fig. \ref{ayppdpi} shows the results for
the transverse analyzing power $A_y$ in the basic input
reaction $pp \rightarrow d\pi^+$ at two energies
close to the energies of the data in $\pi^+$ absorption on
$^3$He.
If the quasifree ansatz is correct and the employed
wave functions are realistic, one would expect a similar degree
of agreement also in the latter reaction.
Please, note that in the calculations of the
$pp \rightarrow d\pi^+$ observables
the $s$-wave rescattering
form factor has been adjusted individually
for each deuteron wave function
to reproduce the depth of the dip at $90^\circ$
as shown by the data at 515 MeV \cite{aprile}
(the cut-off mass used is $\Lambda = 3m_\pi$ for CD Bonn, $4m_\pi$ for
Paris and $5m_\pi$ for Reid).
It is interesting to observe that there are differences between
the results at the higher energy - even after this fitting -
and that the data may favour the newer wave functions
based on the Paris and CD Bonn potentials vs. the older Reid
soft core model. Without the above described fixing at 515 MeV the
differences with different wave functions would be even
larger. Also, we want to point out that the magnitude of $A_y$ at the
higher energy is strongly correlated with the $D$-state probability
in the employed deuteron wave function, 
with $A_y$ becoming larger with increasing $P_D$.
None of the wave functions is able to reproduce correctly
the dip in the data at 800 MeV.
In these calculations (as in those for absorption, that will be
presented in the next section)
all the partial wave amplitudes up to $J=5$ were included,
which was found sufficient also at 800 MeV.

In the present context the above two-nucleon mechanisms are
embedded in
$^3$He for which we use the parameterization of the full
antisymmetric wave function \cite{param}.
Since then absorption on any pair
should give the same results as on the others we can assume the coordinate
$r$ (e.g. $r_{12}$) to be the active one and particle 3 to be
the spectator. With the above described parameterization
the wave function $\Psi^\nu(r,\rho) = \sum_\lambda v_\lambda^\nu(r)\,
w_\lambda^\nu(\rho)$ would give, for example, for the cross section the
result
\begin{equation}
\frac{d\sigma}{d\Omega} = {\rm Tr} \sum_{\lambda\lambda'\nu\nu'}\,
 (M_{\lambda'}^{\nu'})^*\, M_\lambda^\nu\,
{\cal W}^{\nu' \nu}_{\lambda' \lambda}\, \delta_{\nu\nu'},
\end{equation}
where $M_\lambda^\nu$ stands for the two-nucleon transition
matrix calculated for the state component $\nu$ and for
a specific term $\lambda$ of the parametrization of the
$3N$ wave function. The trace is over spin orientations. Here
the spectator effect has reduced to mere overlap integrals
\begin{equation}
{\cal W}^{\nu ' \nu}_{\lambda ' \lambda} =
\int d\rho\, w_{\lambda '}^{\nu '}(\rho)\, w_\lambda^{\nu}(\rho) .
\end{equation}
Similar expressions with different $\nu$ assignments hold
also for other spin dependent observables.

\section{Results}

\begin{figure}[t!]
\epsfig{figure=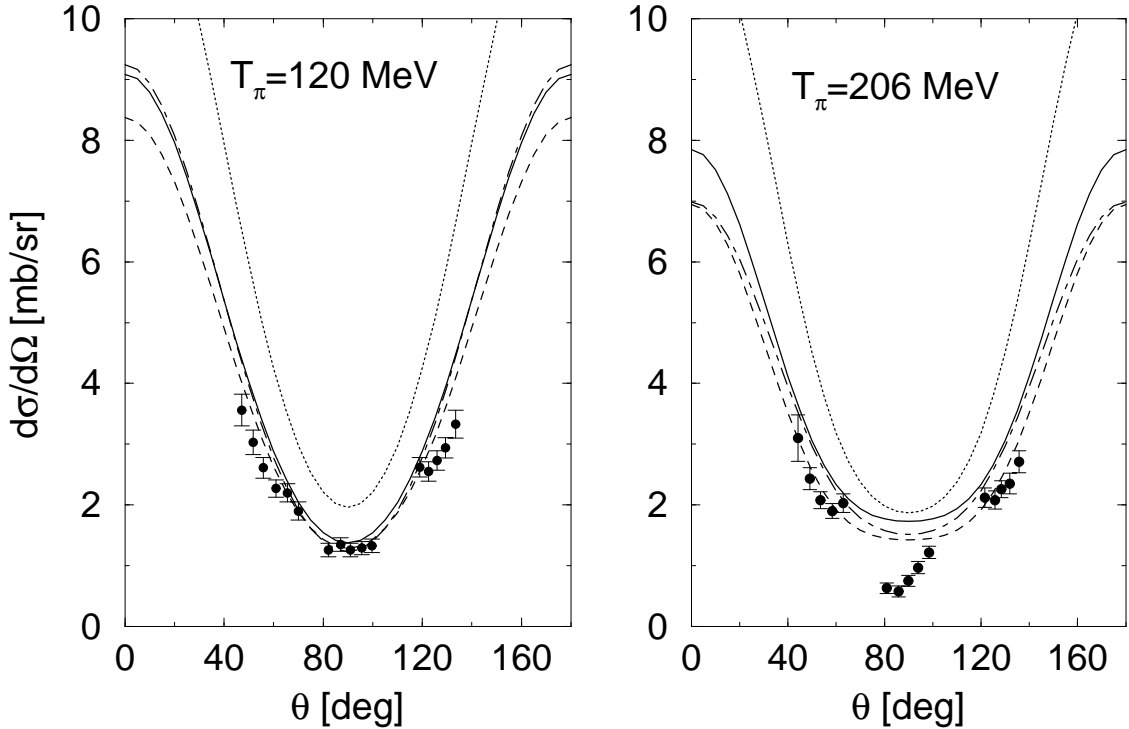,width=10cm,angle=-90}
\caption{The bound state wave function dependence of the
differential absorption cross section at $T_\pi$ = 120 and 206 MeV.
Solid: CD Bonn; dashed: Paris; dash-dotted: single Faddeev amplitude
(normalized to one)
used instead of the fully antisymmetric wave function; dotted: old
result of Ref. \protect\cite{quasid} based on the correlation function.
The data are from Ref. \protect\cite{weber}.  \label{modelx}}
\end{figure}

\begin{figure}[t!]
\epsfig{figure=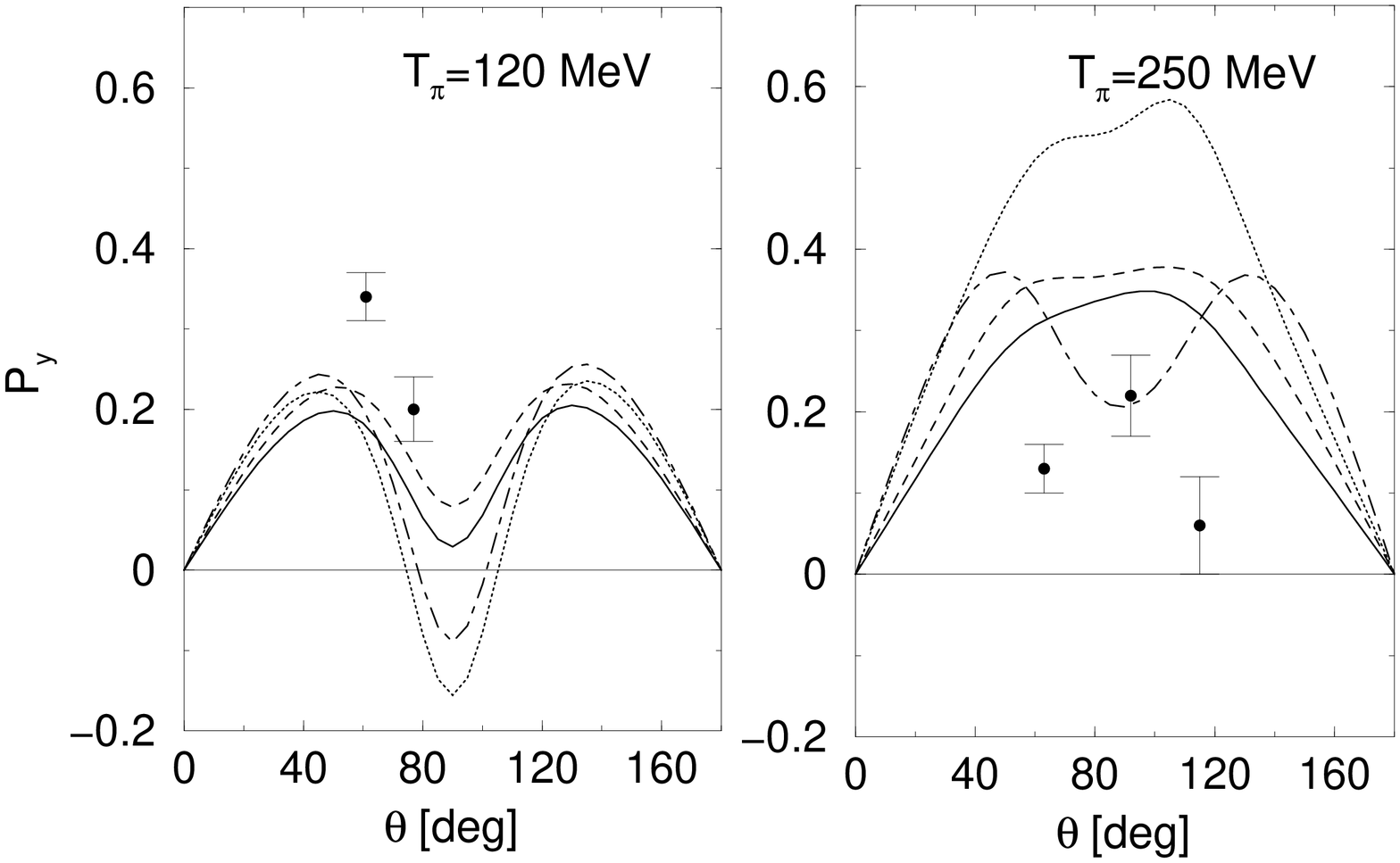,width=10cm,angle=-90}
\caption{The bound state wave function dependence of the
polarization $P_y$ at $T_\pi$ = 120 and 250 MeV.
Notation as in Fig. \protect\ref{modelx}; the data
from Ref. \protect\cite{maytalbeck}. \label{modelp} }
\end{figure}

In Figs. \ref{modelx} and \ref{modelp} we show the differential
cross sections and proton polarizations for pion absorption on
$^3$He at two energies for various trinucleon wave functions 
using the two-term separable fits given in Ref. \cite{param}.
Our particular aim is to study the behaviour of the proton
polarization for different wave functions
and compare the results with the data of Ref. \cite{maytalbeck}.
In addition to the quasideuteron, in the present calculations also
absorption on the $^1S_0\; np$ pair is included, although
without the heavy meson exchange (Fig. \ref{mechs}e)
its influence has been found to be very small
\cite{quasids}. However, the heavy meson exchange
enhances its contribution in $s$-wave absorption of negative
pions on $^3$He \cite{impact}.

We find that, even after adjusting the amplitude 
for the basic reaction $pp\rightarrow d\pi^+$
in order to get agreement with the analyzing power $A_y$ at 515 MeV,
the CD Bonn (solid) and Paris (long dashes)
potentials give somewhat different results
for both the absorption cross section and polarization on $^3$He (although
the angular distributions are rather similar).
The difference in the total cross section is about 10\%, but
without the adjustment it would be 20--30\%, i.e. comparable
to the spread obtained in Ref. \cite{horod} for the
two-nucleon reaction $pp\rightarrow d\pi^+$ using several different
deuteron wave functions.
%

With the new trinucleon bound state wave functions we find a much
better agreement with data than in earlier studies\cite{quasid}. There 
is still a slight overestimation (up to 30\%), but also the data seem 
to have some problems at least at 206 MeV. The overestimation might
be traced back to at least
four possible unaccounted sources contributing. Firstly, inclusion
of nonabsorptive break-up of the $^3$He should decrease other modes
of inelasticity. Secondly, the normalization of the five most
important partial waves to unity in Ref. \cite{param} may lead to
a few per cent overestimation, since higher partial wave components
in the bound state would not contribute much to the total cross
section. Also the free three-nucleon final state, treated at
the two-nucleon level, does not take into account possible
normalization effect from non-orthogonality of different
three-nucleon permutations. Furthermore, the minor effect of the Coulomb
repulsion has been neglected.

Contrary to Ref. \cite{maytalbeck} the qualitative shape
of the polarization
can now be roughly reproduced with the
new bound-state wave functions (solid and dashed curves)
also at the higher energy 250 MeV.
The slight minimum in the old calculation has all but vanished.
This result may be related to the smaller deuteron $D$-state probability,
as in the earlier calculations neglecting this contribution
reproduced the high-energy data \cite{maytalbeck}. (However, then
the 120 MeV result was unacceptable.) In Ref. \cite{param} the
momentum representation of the $D$-state wave function of the
pair was seen to depend strongly also on the spectator momentum.

The results discussed above were all obtained with
the parameterization of the full antisymmetric wave function.
The result with a single Faddeev amplitude (normalized to one;
see Eqs.~(\ref{single}) and (\ref{total})) is given as the
dash-dotted curve. This
wave function has a significantly different short-range behaviour
including a node at about 0.3 fm and is similar to the functions
provided in Ref. \cite{hgs} also used in the calculations
of negative pion absorption on $pp$ pairs \cite{piminus}.
It is a very striking result that the single Faddeev amplitude
gives a qualitatively unacceptable polarization.
Furthermore, the dotted curve shows the original
"best" result from Ref. \cite{quasid} where the bound state wave
function was based on a correlation function of protons in $^3$He
given in Ref. \cite{friar}. The longer range of the full
antisymmetrized wave function obtained in Ref. \cite{param} is
clearly reflected in the results, most directly in the decrease of the
cross section.

\begin{figure}[t!]
\epsfig{figure=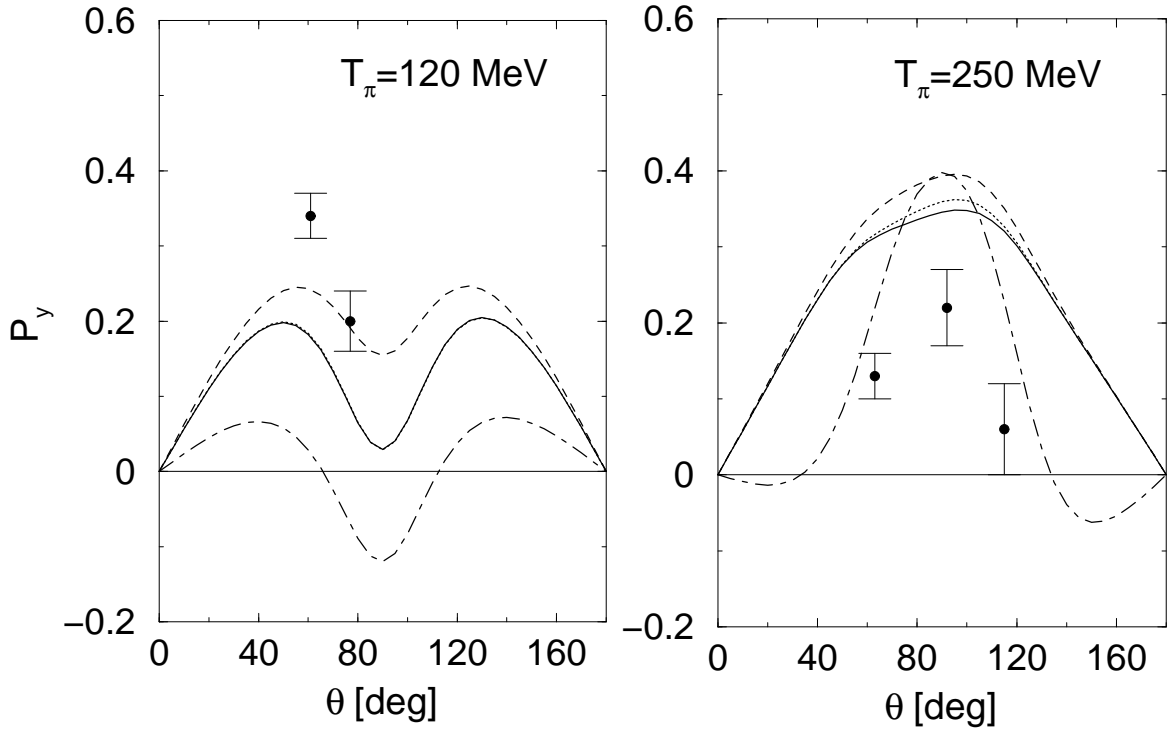,width=10cm,angle=-90}
\caption{The effect of different interaction components on
the polarization $P_y$ at 120 and 250 MeV. Solid: CD Bonn
full result as in Fig. \protect\ref{modelp}; dotted:
absorption on the $^1S_0$ pair neglected; dashed: heavy meson
exchange mechanism neglected; dash-dotted: the $D$-state of the
quasideuteron omitted. The data
from Ref. \protect\cite{maytalbeck}. \label{modeldep} }
\end{figure}

Another method to study the model dependence is to switch on
and off different interaction components. Corresponding
results are shown in Fig. \ref{modeldep} for the polarization.
The solid curve is the same as in Fig.
\ref{modelp} (CD Bonn), while in the dotted curve absorption on the
$^1S_0$ $np$ pair is neglected. In spite of the heavy meson exchange
(Fig. \ref{mechs}e) enhancing this contribution,
the effect is still very small as was
found earlier in Ref. \cite{quasids} without this exchange.
The present results include also the heavy meson exchange effect
in absorption on the quasideuteron. In the dashed line this
heavy meson exchange is neglected. Due to the
more condensed bound state wave function, its effect is somewhat
larger than in  $pp\rightarrow d\pi^+$,
but still does not affect the results qualitatively.

Finally the dash-dotted curve
shows results where the quasideuteron $^3D_1$ component has been
completely neglected. Now the shape of the polarization at 250 MeV
is well reproduced but then the results at the lower energy are 
strongly at variance with the data 
-- even qualitatively, as was also seen in Ref. \cite{maytalbeck}
for a wave function based on an Argonne potential calculation.
Although there is some uncertainty in the $D$-state
component (in particular its probability is unknown), it is
obvious that it cannot be made small enough for a perfect
agreement with the higher energy data.

\section{Conclusion}

In this paper we have employed new and improved
parametrizations of three-nucleon wave functions, obtained 
from Faddeev calculations with realistic $NN$ models, 
with the aim of investigating the crucial
observables of quasifree pion absorption on two nucleons in
a three-nucleon environment.
The most essential new points were the use of a total
antisymmetrized wave function of the target and a nonseparable
fitted form of its wave function to allow for correlations between
the two relative canonical coordinates (or momenta).
Both were seen to have an effect.

Although an essential improvement is found, these results confirm
to some extent the earlier result (which used less
sophisticated bound state wave functions) that conventional
two-nucleon calculations cannot be easily accommodated with
the polarization data of \cite{maytalbeck,aclander}. In these
earlier works a somewhat poorer description was obtained for pair
wave functions with a sizable $^3D_1$ component (quasideuteron).
However, if the effect of the $D$ state was neglected, the
high-energy polarization could be reproduced at the expense of the
agreement at 120 MeV as seen in Fig. \ref{modeldep}. The present
improvement over the old results may be partly seen as a compromise
with somewhat smaller $D$-wave components in the new
deuterons and quasideuterons. However, it is hardly realistic
to assume an arbitrarily low $D$-state probability
only in order to agree with the higher energy
results. In spite of these deviations, considering the quality
of the agreement in the basic $pp\rightarrow d\pi^+$ reaction
and the uncertainties of the data (only altogether five points at
two distant energies) it may be possible to regard the present
agreement achieved with new, more realistic
trinucleon wave functions as qualitatively acceptable.

For further improvements one might have to consider
either some energy-dependent mechanism yet not included
(and possibly not necessary in two-nucleon reactions) or admit
some influence, probably active participation, of the
spectator even in the quasifree kinematics of the conjugate angles.
An argument for this may be found indirectly from a comparison of
the true two-body and quasi-two-body results. In our model
calculations the characteristic basic structure of the
{\it calculated} polarization
in $pp\rightarrow d\pi^+$ (see Fig. \ref{ayppdpi})
was carried over to pion absorption on $^3$He for a variety
of $3N$
bound state wave functions. Some success in the absorption
reaction was achieved perhaps, because the dip in the two-nucleon
data at 800 MeV was not reproduced by the basic $pp\rightarrow d\pi^+$
amplitude. Indeed,
one can deduce that, if the assumption of quasifree
mechanisms is valid {\it and} the abundant two-nucleon data for
$A_y$ at 800 MeV are used, then it is hard to understand why
the dip structure should not be found also in the polarization of
pion absorption on quasideuterons.
Since the existing data at 250 MeV clearly do not indicate such
a dip, they lend support for the need of other mechanisms.

The latter possibility would make nuclear physics with mesonic
inelasticities even more involved than previously believed.
On the other hand, this could be a tool to study the effect
of the "medium" on pionic inelasticities. Therefore,
it would be desirable to have more data to investigate in
detail the development of $P_y(\theta)$ for energies
intermediate to 120 and 250 MeV and also to confirm the
structural difference seen in the data at 250 MeV.
Also it may be
useful to apply the new wave functions in $\pi^-$ absorption on the
singlet $pp$ pair in $^3$He or $\pi^+$ on the $nn$ pair in
a triton \cite{future}. In this case the two-body
absorption is strongly suppressed making the possible (but also
small) $3N$ background more visible.

\begin{acknowledgments}
Financial support for this work was provided in part by the
international exchange program between DAAD (Germany, project no.
313-SF-PPP-pz) and the Academy of Finland (project no. 41926).
\end{acknowledgments}

\end{document}